# Análisis jurídico de la discriminación algorítmica en los procesos de selección laboral

Natalia Ramírez-Bustamante•
Andrés Páez°

RESUMEN

El uso de sistemas de *machine learning* en los procesos de selección laboral ha sido de gran utilidad para agilizarlos y volverlos más eficientes, pero al mismo tiempo ha generado problemas en términos de equidad, confiabilidad y transparencia. En este artículo comenzamos explicando los diferentes usos de la Inteligencia Artificial en los procesos de selección laboral en Estados Unidos. Presentamos los sesgos sexuales y raciales que han sido detectados en algunos de ellos y explicamos los obstáculos jurídicos y prácticos para su detección y análisis. Un posible remedio jurídico a la discriminación algorítmica es el régimen de impacto diferencial sistémico utilizado en Estados Unidos. Usamos algunas características de este régimen para señalar los vacíos jurídicos frente a la discriminación en el acceso al empleo en el ámbito del derecho laboral colombiano, y ofrecemos algunas características generales que podría tener un régimen de esa naturaleza.

## 1. Introducción

Durante los últimos años, las grandes compañías del mundo han empezado a utilizar un software conocido como ATS, por sus siglas en inglés (*Applicant Tracking System*), en los procesos de selección de personal. Los ATS son usados para recolectar, clasificar y calificar las solicitudes de trabajo que reciben. En varias etapas del proceso de selección, las empresas utilizan sistemas de *machine learning*, o aprendizaje automático. Los sistemas son entrenados para encontrar patrones en los datos de los aspirantes con el fin de tomar decisiones de contratación laboral. Existen varios algoritmos diferentes en el mercado y cada uno tiene características particulares, pero en términos generales todos buscan detectar en los candidatos rasgos que permitan predecir su buen desempeño laboral. Es justamente en la búsqueda de estas señales de éxito que surgen sesgos involuntarios, algunos de los cuales pueden ser discriminatorios.

---

• Profesora Asistente de la Facultad de Derecho de la Universidad de los Andes y directora del semillero de investigación Trabajo y Derecho de la misma universidad.
° Profesor Asociado del Departamento de Filosofía de la Universidad de los Andes, director de Philogica: Grupo de Investigación en Lógica, Epistemología y Filosofía de la Ciencia, y miembro del Centro de Investigación y Formación en Inteligencia Artificial (CinfonIA).

Aunque el uso de estos mecanismos en América Latina y particularmente en Colombia es todavía restringido, algunas multinacionales con presencia en la región han empezado a utilizar estas herramientas en sus procesos de selección de personal. Sin embargo, el sistema jurídico colombiano no cuenta con herramientas eficaces para combatir la discriminación resultante de estos procesos. Este artículo llama la atención sobre este vacío normativo y sus implicaciones, así como sobre la necesidad de regular los procedimientos previos a la contratación de personal.

Este artículo está dividido en seis partes que siguen a esta introducción. En la segunda parte, describimos el funcionamiento de algunos de los algoritmos de selección de personal más conocidos. En la tercera parte, indicamos los problemas de discriminación que ocurren en la utilización de estos mecanismos automatizados. En la cuarta parte exploramos los obstáculos prácticos y teóricos para la investigación de la discriminación algorítmica. En la quinta parte, describimos el régimen de impacto diferencial sistémico estadounidense, así como sus limitaciones para enfrentar la investigación y sanción de la discriminación en el empleo. En la sexta parte llamamos la atención sobre el vacío normativo existente en nuestro ordenamiento a falta de un estatuto antidiscriminación, y para concluir ofrecemos algunas características generales que podría tener un régimen de esa naturaleza.

**2. Los algoritmos de selección de personal**

Los ATS son utilizados por los departamentos de recursos humanos de las grandes compañías para procesar un gran número de solicitudes de trabajo. En 2018, el 98% de las compañías en el ranking *Fortune 500* utilizaban algún tipo de ATS[1]. Los ATS permiten organizar y rastrear de manera centralizada todo el proceso de selección, desde la publicación y difusión de las ofertas laborales, pasando por la búsqueda de candidatos en sitios que consolidan perfiles laborales, hasta el proceso de evaluación y contratación. Dado el volumen de hojas de vida que reciben diariamente las empresas, los ATS utilizan herramientas de Inteligencia Artificial (IA) en varias etapas del proceso. Las ofertas de empleo son promocionadas en internet utilizando el software de plataformas publicitarias virtuales como Google Ads o los algoritmos de las redes sociales. Posteriormente se hace una revisión inicial de las hojas de vida recibidas con lectores automáticos de texto que usan palabras clave seleccionadas por el empleador y se lleva a cabo una búsqueda de información adicional sobre el solicitante en internet. Existen compañías como Fama Technologies[2] especializadas en buscar publicaciones racistas, sexistas o violentas de los candidatos en redes sociales para descartarlos. Cada vez es más frecuente que el proceso de evaluación incluya una entrevista con un sistema de IA. Los candidatos son entrevistados utilizando chatbots para extraer información adicional y depurar la lista de finalistas. Entre los chatbots más conocidos están *Mya*, *Olivia*, *Myra* e *Yva*, curiosamente todos nombres femeninos[3]. Los bots también les permiten a los aspirantes estar al tanto del estado de su proceso de selección. Los programas de evaluación más sofisticados, como los utilizados por la compañía HireVue[4], analizan los patrones de habla y las expresiones faciales de los aspirantes en videos grabados mientras responden una serie de preguntas. Según la compañía, cada minuto de video les proporciona más de un millón de datos que son analizados por sistemas de *machine learning* para detectar características como inteligencia emocional y aptitudes sociales. Además, los sistemas de HireVue examinan los resultados de pruebas cognitivas y neurológicas tomadas por los candidatos, algunas de las cuales tienen el formato de un videojuego. De acuerdo con HireVue, el resultado de estos análisis les proporciona a las empresas información acerca del estilo de trabajo del candidato, su capacidad para trabajar en grupo, su personalidad y sus habilidades cognitivas generales, habilidades blandas que no pueden ser deducidas fácilmente de su historia laboral.

---

[1] Jon Shields, "Over 98% of Fortune 500 Companies Use Applicant Tracking Systems (ATS)", *Jobscan*, junio 20, 2018, https://www.jobscan.co/blog/fortune-500-use-applicant-tracking-systems/.

    Además de ser eficientes y efectivos en el procesamiento de una gran cantidad de información, los ATS también han sido promocionados por las ventajas cualitativas que

---

[2] www.famatechnologies.com

[3] Los detalles de cada chatbot pueden ser consultados en las páginas de sus fabricantes: hiremya.com, www.olivia.ai, www.myralabs.com, yva.ai.

[4] www.hirevue.com

tienen frente a la selección laboral hecha por humanos a partir del examen de hojas de vida, cartas de referencia y entrevistas presenciales o por videoconferencia. Entre estas ventajas se encuentran la eliminación de las inconsistencias que ocurren frecuentemente entre una entrevista y otra debido a las diferencias en las preguntas que se les formulan a los candidatos. También elimina la distorsión generada por el contraste entre un candidato bueno seguido de uno promedio, o al revés. También se promociona su capacidad para evitar sesgos sociales como el uso de estereotipos raciales, sexuales o económicos; sesgos no verbales, como los ademanes y gestos de las personas, que pueden mejorar o empeorar el efecto de sus respuestas; y los sesgos cognitivos, como el sesgo de negatividad (darle más peso a los aspectos negativos que a los positivos en una hoja de vida) y el efecto halo, que es la tendencia a dejarnos impresionar por características de las personas que no son relevantes para el trabajo en cuestión, como el uso de una corbata o una mancha en la camisa.

## 3. Casos de discriminación algorítmica

La discriminación algorítmica de que pueden ser objeto los aspirantes a un empleo comienza mucho antes de someter sus hojas de vida a un ATS. Existen varios estudios que muestran cómo los algoritmos de Google Ads y de Facebook seleccionan la población a la cual van dirigidas las ofertas laborales de acuerdo con criterios no equitativos. No se trata de una estrategia discriminatoria de los anunciantes, que podrían escoger de manera sesgada al público al cual va dirigida una oferta. Se trata, más bien, de un sesgo generado por el algoritmo que distribuye las ofertas laborales y decide a cuáles grupos poblacionales se las muestra. Por ejemplo, una serie de 21 experimentos en la Universidad Carnegie Mellon utilizando 1000 perfiles ficticios, repartidos equitativamente entre hombres y mujeres, mostró que Google Ads tendía a mostrarles los trabajos mejor remunerados a más hombres que mujeres. Un trabajo en particular, que pagaba más de US$200,000, fue mostrado 1800 veces a los hombres y solo 300 veces a las mujeres[5]. Aunque los investigadores pudieron

---

[5] Amit Datta, Michael Carl Tschantz y Anupam Datta, "Automated Experiments on Ad privacy Settings: A Tale of Opacity, Choice, and Discrimination", *Proceedings on Privacy Enhancing Technologies* 2015, no. 1 (2015): 92.

identificar estas discrepancias, no las pudieron explicar porque ocurren dentro de la caja negra del algoritmo de Google Ads[6].

También existe evidencia de que el algoritmo de publicidad de Facebook distribuye de manera sesgada las ofertas de trabajo pautadas en la plataforma publicitaria. En su página, Facebook tiene una política explícita acerca de la discriminación en los anuncios laborales:

> Los anuncios laborales no pueden discriminar de manera ilegal a los aspirantes con base en ninguna característica protegida, incluyendo, pero no limitada a, raza, etnicidad, color, origen nacional, religión, edad, sexo, orientación sexual, identidad de género, estatus familiar, discapacidad, condiciones médicas o genéticas o cualquier otra protegida por leyes federales, estatales o locales[7].

Facebook no vigila directamente el cumplimiento de esta política, sino que apela a la autorregulación de los anunciantes. Sin embargo, un estudio de Ali et al.[8] mostró que incluso cuando los anunciantes tienen la intención de cumplir con esta política y distribuir la oferta de manera equitativa, tanto el contenido del anuncio como el presupuesto del anunciante generan sesgos raciales y sexuales en la forma en que el algoritmo de Facebook decide a quién mostrar el anuncio. El algoritmo toma decisiones acerca de la "relevancia" de la oferta para diferentes grupos de usuarios a partir de su contenido, incluyendo las imágenes que aparecen en el anuncio. A partir de un extenso registro de los historiales de uso, y de la frecuencia con que abren anuncios con ciertos tipos de contenido, el algoritmo aprende los hábitos y preferencias de los usuarios de la red social y decide cuáles ofertas mostrarles a esas personas. Los hábitos y preferencias de los usuarios generan sesgos que no fueron planeados por los anunciantes, y estos a su vez están determinados por factores como su género, su raza y su estatus socioeconómico. En el experimento llevado a cabo por los autores en Estados Unidos, se pautaron tres anuncios diferentes en Facebook, con la especificación de que debían ser dirigidos exactamente el mismo público. Pero el algoritmo de distribución tenía sus propias ideas al respecto: el primer anuncio, que ofrecía una posición en la industria maderera, le llegó a una población 72% blanca y 90% masculina; en el segundo, que ofrecía posiciones en la caja de un supermercado, el 85% de los receptores fueron mujeres; y en el tercero, que ofrecía trabajo en una compañía de taxis, el 75% de los receptores fueron afroamericanos[9].

---

[6] Latanya Sweeney, "Discrimination in Online Ad Delivery", *Queue* 11, no. 3 (2013): 10-29, hace un análisis de los sesgos raciales de los anuncios asociados a los nombres de pila usados en el motor de búsqueda de Google.

[7] Facebook (2020), "Advertising Policies: 3. Discriminatory Practices", https://www.facebook.com/policies/ads/prohibited_content/discriminatory_practices#

[8] Muhammad Ali, Piotr Sapiezynski, Miranda Bogen, Aleksandra Korolova, Alan Mislove y Aaron Rieke, "Discrimination through Optimization: How Facebook's Ad Delivery Can Lead to Skewed Outcomes", *arXiv preprint arXiv:1904.02095* (2019): 1.

Además, los sistemas de distribución de anuncios identifican grupos demográficos "valiosos", es decir, aquellos que tienen una alta probabilidad de ver el anuncio y hacer una compra, o en este caso, enviar una hoja de vida. El sistema no distribuye el anuncio entre los usuarios valiosos a menos que los anunciantes tengan un presupuesto holgado que les permita acceder a ellos. Si este grupo demográfico valioso es además un grupo privilegiado, el sesgo algorítmico termina siendo discriminatorio. De esta forma, tanto el contenido de los anuncios, como el presupuesto invertido por el anunciante, termina generando sesgos en su distribución, muchos de los cuales son discriminatorios[10].

Otra fuente notable de discriminación algorítmica ocurre cuando los empleadores buscan candidatos en bases de datos laborales como Indeed, Monster o CareerBuilder. Chen et al.[11] examinaron los algoritmos utilizados por los motores de búsqueda de estos tres sitios, los cuales permiten filtrar los resultados y utilizar palabras clave. Como todos los motores de búsqueda, los algoritmos ordenan a los aspirantes, mostrando primero a los que obtienen el mejor puntaje. Los autores descubrieron que los ordenamientos de los motores de búsqueda tenían sesgos de género. El estudio se basó en 35 ofertas de trabajo en 20 ciudades de Estados Unidos, y la base de datos sobre la cual se hizo la búsqueda tenía 855,000 personas. Los autores estudiaron dos tipos de discriminación. Inicialmente, examinaron si los motores de búsqueda exhibían *justicia individual*, es decir, si los candidatos con calificaciones similares eran ordenados en posiciones similares. Posteriormente examinaron si los motores de búsqueda exhibían *justicia grupal,* es decir, si la distribución de hombres y mujeres que arrojaba la búsqueda correspondía a la distribución real en la población. El resultado indica que los tres algoritmos muestran un sesgo de género leve en el caso de la justicia individual, pero un sesgo significativo en el caso de la justicia grupal. En 12 de las 35 ofertas de trabajo los algoritmos favorecían consistentemente a los candidatos masculinos[12].

---

[9] Ali et al., "Discrimination through Optimization", 2.
[10] A raíz de una denuncia de Pro Publica, Facebook fue demandada por discriminación laboral y residencial en 2016 (*Mobley v. Facebook.* 5:16-cv-06440 (N.D. Cal.). 2016). La justificación de la demanda era que Facebook les permitía a los anunciantes decidir por "afinidad étnica" el público de sus anuncios. Al contratar los servicios de Facebook, los anunciantes podían excluir determinados grupos étnicos. El caso fue conciliado en 2019 y Facebook eliminó la posibilidad de decidir la audiencia de los anuncios laborales, residenciales y crediticios utilizando categorías como edad, género, etnicidad o código postal. Este no es un caso de discriminación algorítmica no intencionada, sino de una estructura que permitía la discriminación directa.
[11] Le Chen, Ruijun Ma, Anikó Hannák y Christo Wilson, "Investigating the Impact of Gender on Rank in Resume Search Engines", en *Annual Conference of the ACM Special Interest Group on Computer Human Interaction* (Montreal, Canadá, abril de 2018), 1-14.

En 2017 la Procuraduría del estado de Illinois había enviado una carta de advertencia a seis bases de datos laborales, incluyendo las tres examinadas en el estudio de Chen et al., tras recibir quejas de discriminación por edad. La Procuraduría encontró que al intentar ingresar una hoja de vida en estas bases de datos, los rangos de fechas en los menús desplegables impedían que candidatos nacidos antes de cierto año pudieran hacerlo[13]. A diferencia de los resultados encontrados por Chen et al., este tipo de discriminación es intencional e inmediatamente detectable.

Uno de los casos más sonados de discriminación algorítmica fue el intento de Amazon de diseñar un algoritmo de selección de hojas de vida. La compañía cuenta actualmente con 840,000 empleados en todo el mundo[14] y el Departamento de Recursos Humanos debe procesar miles de solicitudes de trabajo cada día. Un algoritmo propio que pudiera facilitar esta tarea hubiera sido de gran ayuda y ahorraría mucho dinero. El propósito del algoritmo era calificar a los candidatos usando el mismo sistema de cinco estrellas que se usa en la página de Amazon para calificar los productos que venden. El equipo técnico creó 500 modelos diferentes adaptados para posiciones y ubicaciones diferentes. No se conocen los números exactos, pero el proyecto fue desechado después de que se descubriera que los modelos tenían un fuerte sesgo de género, pues habían sido entrenados con las hojas de vida de trabajadores exitosos de los últimos 10 años, y esa base de datos contenía una inequidad de género muy común en la industria de la tecnología. El 63% de los empleados durante ese periodo habían sido hombres, cifra que subía a 75% cuando se consideraban solo los cargos ejecutivos. En una fase inicial, cuando el algoritmo encontraba una referencia a una actividad que usara la palabra "mujer", como en "club de tenis de mujeres", la hoja de vida era peor calificada. Los analistas de Amazon corrigieron ese detalle, pero el algoritmo encontraba otras formas de detectar el género de los aspirantes. Por ejemplo, al detectar las palabras "ejecuté" y "capturé", que son usadas típicamente por ingenieros hombres, el algoritmo mejoraba el puntaje del aspirante[15]. Tras el fiasco, hoy en día Amazon usa una forma más rudimentaria de ATS.

---

[12] Se ha encontrado un sesgo similar en TaskRabbit y Fiverr, dos bases de datos que ofrecen los servicios de trabajadores por horas (*freelance*). Véase Anikó Hannák, Claudia Wagner, David Garcia, Alan Mislove, Markus Strohmaier y Christo Wilson, "Bias in Online Freelance Marketplaces: Evidence from Taskrabbit and Fiverr", en *Proceedings of the 2017 ACM Conference on Computer Supported Cooperative Work and Social Computing* (2017), 1914-1933.

[13] Illinois Attorney General, "Madigan Probes National Job Search Sites over Potential Age Discrimination", *Illinois Attorney General Press Release,* March 2, 2017, http://www.illinoisattorneygeneral.gov/pressroom/2017_03/20170302.html.

[14] Andria Cheng, "Amazon's Coronavirus Plan to Test all 840,000 Employees May Pressure other Companies to Follow Suit", *Forbes*, abril 16, 2020, https://www.forbes.com/sites/andriacheng/2020/ 04/16/amazons-plan-to-test-all-employees-on-coronavirus-may-up-the-ante-for-other-companies/#b4e0a0736cb1.

El último caso de discriminación algorítmica que presentaremos es el de HireVue, la compañía especializada en el análisis facial de los aspirantes a un empleo a través de videos y en el uso de pruebas con formato de videojuego. En 2019 el Centro para la Privacidad de la Información Electrónica (*Electronic Privacy Information Center*-EPIC), un centro de investigación de interés público, le pidió a la Comisión de Comercio Federal de los Estados Unidos que investigara a HireVue. Una de las quejas tenía que ver con el uso de videojuegos en el proceso de selección, los cuales ponen a los aspirantes de más edad en desventaja. Pero el elemento más relevante de la acusación de EPIC tenía que ver con la forma en que el algoritmo predictivo evalúa los videos de los aspirantes. Según HireVue, el 29% de la calificación de un aspirante se desprende de sus movimientos faciales[16]. La preocupación de EPIC es que el algoritmo discrimina en contra de personas con depresión, algún grado de autismo, timidez extrema, o que no son hablantes nativos de inglés. Según EPIC, el algoritmo de HireVue no cumple con el requisito estipulado por la Comisión de Comercio Federal según el cual debe tener una "base razonable" para las conclusiones a las que llega. En efecto, existen serias dudas en la comunidad científica acerca de la validez de inferir emociones u otros estados mentales a partir de las expresiones faciales[17]. Además, como el algoritmo es secreto y los aspirantes ni siquiera reciben un puntaje, la compañía estaría violando los principios de transparencia de la OECD[18] sobre el uso de la IA[19].

---

[15] Jeffrey Dastin, "Amazon Scraps Secret AI Recruiting Tool that Showed Bias against Women", *Reuters*, octubre 9, 2018, https://www.reuters.com/article/us-amazon-com-jobs-automation-insight/amazon-scraps-secret-ai-recruiting-tool-that-showed-bias-against-women-idUSKCN1MK08G.

[16] Drew Harwell, "A Face-Scanning Algorithm Increasingly Decides whether You Deserve Job", *The Washington Post*, octubre 22, 2019, https://www.washingtonpost.com/technology/2019/10/22/ai-hiring-face-scanning-algorithm-increasingly-decides-whether-you-deserve-job/.

## 4. Obstáculos para la investigación de la discriminación algorítmica

Existen al menos cinco obstáculos para la investigación de la discriminación algorítmica. El primero tiene que ver con el hecho de que la estructura del algoritmo y los datos de entrenamiento son secretos industriales de acceso restringido. Pero incluso si se tuviera acceso a los detalles del algoritmo y a los datos de entrenamiento, surgen dos obstáculos más. Por una parte, los algoritmos de *machine learning* son cajas negras que no permiten una explicación completa de un resultado discriminatorio. Por otra parte, es muy fácil enmascarar una intención discriminatoria utilizando etiquetas "proxy", es decir, utilizando características que no se refieren directamente a sexo, raza o estatus socioeconómico, pero que las reemplazan funcionalmente. El cuarto obstáculo es de orden puramente jurídico y ocurre principalmente en el contexto estadounidense. Los dueños de los ATS y de las páginas web enfocadas en el mercado laboral han intentado imponer límites jurídicos al examen con fines investigativos de la información contenida en ellos, así esta sea de acceso público. En particular, han intentado obstaculizar jurídicamente el trabajo de periodistas e investigadores que intenten buscar tendencias discriminatorias en sus datos y algoritmos. En los últimos dos años las cortes se han pronunciado para debilitar estos obstáculos, pero no han sido eliminados del todo. Finalmente, el quinto obstáculo es de orden teórico. La discriminación algorítmica se puede entender como un fenómeno individual o como uno grupal, es decir, el algoritmo puede tratar de manera inequitativa a dos individuos con calificaciones semejantes, o puede discriminar a un grupo entero de la población. El problema reside en que los algoritmos pueden discriminar en un sentido, pero no en el otro, como vimos en el experimento de Chen et al. Así, puede surgir un conflicto en la implementación simultánea de ambos ideales de equidad. Examinaremos cada uno de estos obstáculos a continuación.

---

[17] Lisa Feldman Barrett, Ralph Adolphs, Stacy Marsella, Aleix M. Martinez y Seth D. Pollak, "Emotional Expressions Reconsidered: Challenges to Inferring Emotion from Human Facial Movements", *Psychological Science in the Public Interest* 20, no. 1 (2019): 1.

[18] OECD, "Recommendations of the Council on AI", *OECD Legal Instrument 0449* (2019).

[19] Otros ejemplos de discriminación algorítmica y de discriminación deliberada en los procesos de contratación laboral en Estados Unidos se pueden consultar en: Solon Barocas y Andrew D. Selbst, "Big Data's Disparate Impact," *California Law Review* 104 (2016): 671, y en Pauline T. Kim, "Data-Driven Discrimination at Work", *William & Mary Law Review* 58 (2016): 857.

Comencemos con los obstáculos que tienen que ver con el entrenamiento y la estructura del algoritmo. Los algoritmos de *machine learning* utilizados por los ATS son entrenados utilizando hojas de vida reales de personas que han desempeñado un trabajo semejante de manera exitosa, pero también se utilizan hojas de vida de personas que no han tenido un rendimiento satisfactorio para que la máquina aprenda cuáles son las características que se deben evitar. Estas constituyen el conjunto de datos, o *dataset*, seleccionado para entrenar el algoritmo. Además, se selecciona un conjunto de atributos de los aspirantes que son juzgados como deseables y no deseables por los empleadores. Estos constituyen las *etiquetas* que son utilizadas para clasificar las hojas de vida. Sin embargo, los investigadores no tenemos acceso ni a las bases de datos de entrenamiento ni a las etiquetas, en gran medida porque las hojas de vida están protegidas por el derecho a la privacidad, pero también porque los criterios utilizados para incluir una hoja de vida en el *dataset* de entrenamiento hacen parte de los secretos industriales que le dan ventaja competitiva al dueño del algoritmo[20]. El problema es que, sin acceso a los datos de entrenamiento y a las etiquetas utilizadas, es imposible determinar si hubo un sesgo intencional en la selección y clasificación de las hojas de vida, si el sesgo fue el resultado de los prejuicios inconscientes de la persona que hizo la selección o, lo que es más probable, si simplemente refleja la estructura discriminatoria del ambiente donde fueron seleccionadas: "Cuando se hace de manera descuidada, la minería de datos puede reproducir patrones existentes de discriminación, heredar los prejuicios de decisiones anteriores, o simplemente reflejar los sesgos ampliamente compartidos que persisten en la sociedad"[21].

---

[20] La protección de este tipo de secretos industriales fue reforzada en 2016 en Estados Unidos a través del Defend Trade Secrets Act (Pub. L. 114-153).

Ahora bien, incluso si contáramos con toda esta información, aún persistirían dos obstáculos. Por una parte, los algoritmos son, tanto teórica como prácticamente, cajas negras. Ni siquiera las compañías involucradas podrían, en caso de ser requeridas, dar una explicación completa de cómo el algoritmo llegó a su decisión. Sin entrar en muchos detalles, podemos decir que la no linearidad y la aleatoriedad de algunos de los procesos que lleva a cabo el algoritmo hacen imposible que un ser humano pueda dar cuenta de cómo fue ejecutada la función para la que fue programado el algoritmo. Es por eso que la investigación de la discriminación algorítmica debe enfocarse en la detección de dicha discriminación en el *output*, y en los posibles sesgos en el *input*, sin poder pretender entender el camino del segundo al primero. Debe limitarse a formar lo que se conoce técnicamente como explicaciones *post-hoc*, en las cuales cualquier variación en el output debe ser explicada a partir de variaciones en el input. Sin embargo, es importante añadir que la generación de explicaciones para las decisiones algorítmicas de cualquier tipo todavía se encuentra en su infancia, y hay un alto riesgo de generar explicaciones espurias de las decisiones del algoritmo. Retornaremos a las implicaciones jurídicas de esta limitación en la próxima sección.

Por otra parte, en muchos casos contar con los datos de entrenamiento y las etiquetas no es prueba suficiente de una intención discriminatoria por la facilidad con la que esta puede ser enmascarada. Las etiquetas utilizadas para entrenar un algoritmo pueden no hacer referencia en absoluto a una clase protegida, y pueden usar en su lugar otros atributos (*proxies*) de los cuales se puede inferir fácilmente la pertenencia a una clase protegida. Por ejemplo, del código postal o de la preferencia musical de una persona se pueden inferir atributos acerca de su etnicidad y estatus socioeconómico[22]. Estos atributos, aparentemente "neutros", son funcionalmente equivalentes a etiquetas abiertamente discriminatorias, y en términos probatorios son indistinguibles de los errores involuntarios que pueden ocurrir en el proceso de minería de datos.

---

[21] Barocas y Selbst, "Big Data's Disparate Impact", 674.
[22] Michal Kosinski, David Stillwell, and Thore Graepel, "Private Traits and Attributes Are Predictable from Digital Records of Human Behavior", *Proceedings of the National Academy of Sciences* 110 (2013): 5802.

En conjunto, estos tres primeros obstáculos demuestran que tanto teórica como prácticamente es muy difícil investigar y probar la discriminación estudiando el *input* de entrenamiento, y la estructura y funcionamiento del algoritmo. Como veremos a continuación, también existen obstáculos a la investigación del *output* del algoritmo.

El cuarto obstáculo es de índole puramente jurídica. Existen impedimentos legales en el ámbito estadounidense para investigar si los algoritmos generan decisiones discriminatorias. Según la interpretación oficial de una sección del Computer Fraud and Abuse Act (CFAA), está prohibido que los investigadores y periodistas pongan a prueba el software de los ATS en busca de discriminación, pues tales acciones constituyen una violación a los términos de uso de las páginas web y pueden dar lugar a la imputación de delitos penales. Dichos términos son fijados de manera arbitraria por cada página y pueden ser modificados en cualquier momento. Prohíben, por ejemplo, crear múltiples usuarios ficticios como los utilizados en el estudio de Datta et al. citado anteriormente. También prohíben proporcionar información ficticia a las páginas, o utilizar métodos automáticos para registrar los resultados de información disponible al público como el resultado de búsquedas en bases de datos. La American Civil Liberties Union demandó esta interpretación oficial del CFAA en 2016 y la demanda fue aceptada preliminarmente en 2018[23]. El caso fue fallado recientemente a favor de la ACLU[24]. Según el juez: "criminalizar las violaciones a los términos de uso conlleva el riesgo de convertir a cada sitio web en su propia jurisdicción criminal y a cada administrador de página web en su propio legislador"[25].

Este caso se une al de *hiQ Labs v. LinkedIn Corp.*[26], que atañe al uso que se le puede dar a la información disponible públicamente en las páginas web. hiQ Labs[27] es una compañía de análisis de datos que utilizó una técnica conocida como barrido de red (*web scraping*) para obtener y analizar información públicamente disponible de las hojas de vida subidas a LinkedIn. Su modelo de negocio es detectar cuáles empleados están más en riesgo de ser tentados por mejores ofertas de la competencia, y venderles esa información a sus empleadores para que emprendan acciones para retenerlos. En 2017 LinkedIn le exigió a hiQ que suspendiera esa práctica, y esta última respondió con una demanda para impedir que LinkedIn pudiera impedir el acceso a sus datos utilizando el Computer Fraud and Abuse Act. La corte falló a favor de hiQ, aduciendo que imponer tales limitaciones era competencia desleal en contra de un competidor directo. LinkedIn ha manifestado su intención de escalar el caso a la Corte Suprema.

---

[23] *Sandvig v. Sessions,* No. 16-1368 (D.D.C. Mar. 30, 2018).
[24] *Sandvig v. Barr*, Civil Action No. 16-1368 (JDB) (D.D.C. Mar. 27, 2020).
[25] Ibid, 21.
[26] *hiQ Labs v. LinkedIn Corp.*, 938 F.3d 985 (9th Cir. 2019).
[27] www.hiqlabs.com

Estas decisiones van a permitir poner en cintura la discriminación algorítmica al abrir las puertas a la investigación académica y periodística. En todo caso, la evidencia que podrá ser utilizada para establecer que un algoritmo es discriminatorio tendrá que seguir siendo obtenida a través de técnicas como la utilizada por hiQ, es decir, buscando patrones en los outputs resultantes de la operación del algoritmo. Esta técnica ha permitido, por ejemplo, encontrar discriminación racial en Airbnb[28] y manipulación de precios en Amazon[29]. Sin embargo, encontrar patrones de discriminación en las decisiones de sistemas de *machine learning* no permite hacer ninguna inferencia acerca de si ésta fue intencionada o no.

Finalmente, el quinto obstáculo para la investigación algorítmica es de índole muy diferente. Si queremos que el uso de la Inteligencia Artificial sea benéfico para la sociedad, debe estar basado en algoritmos justos. Sin embargo, hay dos formas de entender qué es un algoritmo justo. Por un lado, el objetivo puede ser la justicia individual, que requiere que aquellos individuos que sean similares en aquellos aspectos evaluados por el algoritmo tengan una distribución probabilística similar en los resultados del ejercicio de evaluación o clasificación. Por otro, el objetivo puede ser la justicia grupal, que requiere un tratamiento igualitario, en promedio, para los diferentes grupos demográficos[30]. Como mencionamos anteriormente, es posible que se cumpla uno de estos objetivos sin que se cumpla el otro. El problema teórico reside en que en algunos casos es imposible que se cumplan los dos simultáneamente. La justicia algorítmica no es sumativa: una colección de algoritmos clasificadores que son justos individualmente puede no arrojar un resultado justo en el sentido grupal cuando son usados como parte de un sistema más amplio, y viceversa[31]. Una implicación de este resultado es que enfocarse en la justicia o injusticia de un algoritmo a nivel individual puede llevarnos a juzgar de manera errónea el sistema del cual hacen parte. Este es un problema muy conocido para los estudiosos del derecho constitucional en el que la justicia de las normas individuales puede no generar un sistema justo en su conjunto, o viceversa. En el caso de la justicia algorítmica nos encontramos con un problema similar que nos obliga a hacer una pausa antes de evaluar positiva o negativamente un algoritmo en términos de justicia y equidad.

---

[28] Benjamin Edelman, Michael Luca y Dan Svirsky, "Racial Discrimination in the Sharing Economy: Evidence from a Field Experiment", *American Economic Journal: Applied Economics* 9, no. 2 (2017): 1.
[29] Julia Angwin y Surya Mattu, "How We Analyzed Amazon's Shopping Algorithm". *ProPublica*, septiembre 20, 2016, https://www.propublica.org/article/how-we-analyzed-amazons-shopping-algorithm.
[30] Cynthia Dwork y Christina Ilvento, "Individual Fairness under Composition", *Proceedings of Fairness, Accountability, Transparency in Machine Learning* (New York: ACM, 2018).

**5. El régimen de Impacto Diferencial Sistémico y sus limitaciones**

Como hemos señalado, la idea, relativamente extendida, de que los procesos de contratación mediados por algoritmos son más objetivos que el juicio humano empieza a erosionarse cuando se reconoce que estas herramientas son diseñadas por personas cuyos sesgos cognitivos y preferencias raciales o sexuales, entre otros, se trasladan al diseño y operación de dichas herramientas[32]. Cuando esto ocurre, la IA sirve para perpetuar las divisiones socioeconómicas y las desigualdades sociales[33]. De hecho, algunos académicos han llamado a la selección de personal a través de algoritmos "culling systems" o sistemas de sacrificio, en los que se utiliza la pertenencia de un candidato a una clase (mujeres, exconvictos, personas mayores, por ejemplo) como información a través de la cual ocurre un descarte automático de esos candidatos[34].

---

[31] Cynthia Dwork, Moritz Hardt, Toniann Pitassi, Omer Reingold y Richard Zemel, "Fairness through Awareness", In *Proceedings of the 3rd Innovations in Theoretical Computer Science Conference* (New York: ACM, 2012): 214-226.

[32] Anupam Chander, "The Racist Algorithm?" *Michigan Law Review* 115, no. 6 (2017): 1041.

[33] Lee Rainie y Janna Anderson, "Code-Dependent: Pros and Cons of the Algorithm Age," *Pew Research Center*, Febrero 8, 2017, https://www.pewresearch.org/internet/2017/02/08/code-dependent-pros-and-cons-of-the-algorithm-age/.

[34] Ifeoma Ajunwa, "Automated Employment Discrimination" *Harvard Law and Technology Journal* 34 (forthcoming).

Alguna de la literatura que se ocupa de estudiar la adaptabilidad de las normas antidiscriminación a la prevención y sanción de la discriminación algorítmica en el acceso al empleo ha llegado a la conclusión de que los mecanismos normativos existentes, incluso en sistemas jurídicos con un alto nivel de sofisticación en el tratamiento de la discriminación en el empleo, son insuficientes. En esta sección nos ocuparemos de exponer el régimen estadounidense de Impacto Diferencial Sistémico (en inglés, *Sistemic Disparate Impact*), que es el régimen antidiscriminación con mayor vocación para investigar y atacar la discriminación algorítmica por su enfoque en los *efectos* discriminatorios de políticas prima facie neutrales, aunque tiene limitaciones importantes[35].

Existe un régimen antidiscriminación adicional en el sistema jurídico estadounidense: el régimen de tratamiento sistémico dispar (*Sistemic Disparate Treatment*). Este régimen tiene dos elementos constitutivos: por un lado, un tratamiento formal distinto a personas similarmente situadas y, por otro lado, una intensión de discriminar[36]. Es justamente el requisito de demostrar una intensión de discriminar lo que obstaculiza, en la gran mayoría de casos, el uso de este régimen. Como vimos en la sección anterior, es muy difícil probar la intención discriminatoria estudiando el *input* de entrenamiento, tanto por la dificultad de acceder a los datos como por la posibilidad de que haya habido un enmascaramiento de estos. Además, con frecuencia la discriminación ocurre inadvertidamente como resultado de sesgos cognitivos inconscientes, lo que limitaría la responsabilidad del empleador bajo este régimen[37]. Por estas, entre otras razones, su uso para los fines de investigar y sancionar la discriminación algorítmica es muy limitado[38].

A partir de 1964, y como resultado del movimiento por los derechos civiles, el Congreso de Estados Unidos adoptó el Título VII de los Derechos Civiles que incluye prohibiciones en contra de la discriminación ampliamente considerada en empleo en las categorías de raza, religión, origen nacional y sexo [39]. El régimen de Impacto Diferencial Sistémico surgió jurisprudencialmente como resultado del caso *Griggs v. Duke Power Co.*[40], que fue resuelto por la Corte Suprema de Justicia en 1971. De acuerdo con lo que se demostró en el caso, la empresa Duke Power Co. incluyó, como requisito para acceder a un empleo en la compañía, que los candidatos hubieran terminado la educación secundaria, así como la aprobación de un examen estandarizado. La implementación de estos requisitos adicionales de entrada tuvo el efecto de restringir el número de ciudadanos afroamericanos que podían ser contratados o trasladados a cargos previamente ocupados por blancos. Para mostrar el impacto de la adopción de esta medida, prima facie neutral, los demandantes demostraron que en el estado de Carolina del Norte, donde estaba ubicada la fábrica, el 34 por ciento de los residentes blancos tenían un título de educación secundaria comparado con sólo el 12 por ciento de los afroamericanos del mismo estado; al tiempo, sólo el 6 por ciento de los afroamericanos lograban obtener el puntaje necesario para superar la prueba, comparado con el 58 por ciento de los blancos. El impacto diferencial de la adopción de estas medidas sobre las posibilidades de acceder a un empleo en la compañía entre blancos y afroamericanos mostraba la violación prima facie del Título VII aún si los motivos por los cuales el empleador los hubiera adoptado no fueran discriminatorios ni tuvieran nada que ver en el hecho que los blancos pasaran las pruebas en una proporción mayor que los afroamericanos.

---

[35] Michael Selmi, "Was Disparate Impact Theory a Mistake?" *UCLA Law Review* 53 (2006): 702-782.
[36] Pauline Kim, "Data Driven Discrimination at Work", *William and Mary Law Review* 58, no. 3 (2017): 857-936.
[37] Samuel R. Bagenstos, "The Structural Turn and the Limits of Antidiscrimination Law", *California Law Review* 94, no. 1 (2006): 1-48.
[38] Barocas and Selbst, "Big Data's Disparate Impact", 701.

A partir de *Griggs*, la Corte Suprema de Justicia de Estados Unidos reconoció que la protección contra la discriminación en el empleo incluía una protección a los trabajadores cuando las políticas del empleador, aún siendo prima facie neutrales, tenían un impacto discriminatorio sobre clases de trabajadores que merecen especial protección. Adicionalmente, la decisión estableció que el tipo de pruebas adoptadas por los empleadores debían estar relacionadas con las capacidades necesarias para el desempeño de la labor, es decir, las pruebas deberían ser diseñadas para predecir la idoneidad del candidato para la labor específica a desarrollar y no intentar evaluar al candidato en aptitudes genéricas[41]. Posteriormente, el mismo régimen ha sido utilizado, por ejemplo, para demostrar el efecto diferencial de la exigencia del cumplimiento de requisitos de peso y talla para excluir a las mujeres de algunos cargos aún cuando dichos requisitos no tengan un ánimo discriminatorio[42].

---

[39] La prohibición de la discriminación incluida en el Título VII abarca toda la relación laboral, desde el proceso de selección hasta la terminación del contrato de trabajo incluyendo entrenamiento en el empleo, ascensos y beneficios asociados a la relación laboral, entre otros.
[40] *Griggs v. Duke Power Co.*, 401 U.S. 424 (1971).

Bajo el régimen de impacto diferencial, los empleados demandantes que alegan verse afectados por las políticas empresariales deben estar en la capacidad de demostrar que las políticas de contratación de los empleadores, independientemente de la intención ("regardless of intent") tienen efectos adversos sobre unos grupos más que sobre otros y que los mismos no pueden ser adecuadamente justificados como cualificaciones necesarias para la labor a desarrollar. En este tipo de casos, los demandantes deben probar que, estadísticamente, las políticas del empleador tienen un efecto discriminatorio que opera en contra de las clases protegidas, tarea que tiene implicaciones que suelen dificultar el éxito de los demandantes.

Como herramienta para facilitar la medición del efecto discriminatorio de una política empresarial de selección de personal, la Comisión para la Igualdad de oportunidades de los Estados Unidos, entidad encargada de la aplicación del Título VII, creó la llamada *regla del 80%*. Aplicando esta regla a la selección de personal, se deben comparar las tasas de selección de los grupos más elegidos (hombres blancos, por ejemplo) con las tasas de selección de los representantes de grupos menos elegidos (mujeres negras, por ejemplo) en un mismo proceso de selección. Si el grupo menos seleccionado tiene una tasa que es menos de una quinta parte, o el 80%, de la tasa del grupo más seleccionado, este resultado indica un impacto adverso para la clase protegida. La cuestión aquí se reduce a una diferencia estadísticamente significativa, aunque no es claro qué tan amplia debe ser[43].

---

[41] En 1991, el Congreso hizo una enmienda al Título VII para incluir expresamente las subreglas adoptadas en *Griggs*.

[42] En el caso *Dothard v Rawlinson* (433 U.S. 321 1977), el estado de Alabama impuso requisitos de talla y peso para acceder a cargos de policía. Dentro del material probatorio del caso, los demandantes demostraron que los porcentajes de mujeres que pudieran cumplir dichos requisitos era muy inferior a la de los hombres: mientras que los requisitos de talla excluían al 33 por ciento de las mujeres y al 2 por ciento de los hombres, los requisitos de peso excluían al 22 por ciento de las mujeres y sólo al 2 por ciento de los hombres. En este caso la Corte señala que aún cuando el establecimiento de estos requisitos no tenga un ánimo discriminatorio, cuando los mismos operan para excluir a los miembros de un grupo deben ser abolidos.

Aun si los demandantes logran probar el impacto diferencial sistémico, el empleador puede librarse de responsabilidad alegando que el criterio de selección utilizado, cuyos efectos adversos se denuncian, está relacionado con el empleo y corresponde a una necesidad propia de la labor a desempeñar (denominado el "business necessity criteria"). En el contexto de la IA esto significa que los empleadores deben estar en la capacidad de demostrar que la variable de interés está relacionada con el empleo independientemente de la herramienta que se utilice para identificarla. Sin embargo, las cortes estadounidenses han admitido explicaciones plausibles de los empleadores, lo que reduce el estándar probatorio para los demandados[44]. Si el empleador es incapaz de demostrar la conexión entre la variable de interés y la labor que desarrollarían los candidatos, entonces esta defensa fracasaría[45]. Parte del problema radica en que los demandantes tienen la carga probatoria de obtener y analizar los datos con los cuales se entrenó el algoritmo, así como de establecer cuáles son los posibles sesgos que están afectando los resultados y que restringen la participación de miembros de las clases protegidas, lo cual resulta tremendamente costoso en varios sentidos. Por un lado, este es un procedimiento que puede tomar mucho tiempo y que sin duda requiere la inversión de recursos económicos, en muchos casos de alto valor para el demandante. En general, el simple obstáculo de no contar con los datos iniciales de los candidatos considerados por el algoritmo impide a los demandantes demostrar el impacto diferencial de la aplicación de un criterio de selección. Adicionalmente, por las razones que exploramos anteriormente y que tienen que ver con propiedad intelectual sobre el diseño de los algoritmos, este tipo de investigación se toparía con serias restricciones de acceso para los demandantes. Para resolver esta barrera, algunos académicos han propuesto que la carga de demostrar la validez del algoritmo y su carácter no sesgado debería corresponder a los demandados[46].

---

[43] Barocas and Selbst, "Big Data's Disparate Impact", 702.
[44] Ibid, 706.
[45] Ibid.
[46] Kim, "Data Driven Discrimination at Work", 921.

Finalmente, en el ir y venir de cargas probatorias que implica el régimen de impacto diferencial, aún si el empleador demandado logra demostrar que el requisito que reduce la participación de un grupo protegido está relacionado con una necesidad del negocio, los demandantes pueden, finalmente, demostrar que existe otro mecanismo de selección alternativo, es decir, otros requisitos o dispositivos de selección que protegerían el interés legítimo del empleador de contar con una mano de obra eficiente y confiable pero sin impacto discriminatorio, pero que el empleador se niega a utilizarlo. Si, por ejemplo, un demandante puede demostrar que el algoritmo parte de supuestos sesgados que afectan la selección de candidatos, el mecanismo alternativo de selección podría consistir en eliminar los sesgos del algoritmo[47]. Sin embargo, también en este caso los demandantes se enfrentan con una barrera procesal difícil de sobrepasar. Dado el carácter de "caja negra" que tienen los algoritmos de *machine learning*, para un demandante puede ser imposible determinar con certeza exactamente cuáles son los factores que el algoritmo está teniendo en cuenta en su proceso de selección; de acuerdo con Barocas y Selbst, "casi por definición, es difícil saber qué características harán del modelo más o menos discriminatorio. En efecto, a menudo es imposible saber qué características faltan porque los expertos en minería de datos no operan con relaciones causales en mente"[48]. En estas condiciones, formular una alternativa menos discriminatoria parece imposible para el demandante.

Dadas sus características operativas, académicas como Ifeoma Ajunwa han afirmado que los mecanismos de IA empleados en la selección de personal obstaculizan tanto como permiten, la discriminación en el acceso al empleo. En particular, la barrera actualmente infranqueable de la propiedad intelectual sobre los algoritmos hace que sea virtualmente imposible para un demandante tener acceso al diseño y operación de los algoritmos de selección y por ello se reducen sus posibilidades de éxito en una demanda bajo el régimen de impacto diferencial sistémico[49]. De otro lado, las empresas no tienen la obligación de guardar

---

[47] Barocas y Selbst, "Big Data's Disparate Impact", 709.
[48] Ibid, 710.
[49] De acuerdo con Rachel Goodman de la American Civil Liberties Union (ACLU), aunque los empresarios que ofrecen los sistemas de IA de selección de personal afirman que sus herramientas son menos sesgadas que el criterio humano, el carácter privado del funcionamiento de los algoritmos hace que sea imposible verificar esa

un registro de los candidatos que aplican a un trabajo, con lo cual los demandantes no contarían con material probatorio suficiente para determinar el universo de candidatos en el *input* versus el resultado de la selección del algoritmo en el *output*. Adicionalmente, la carga probatoria que enfrentan los demandantes en estos casos implica un alto contenido de análisis estadístico de datos que están en posesión de los demandados, lo que hace del proceso altamente arduo y complejo[50]. En respuesta a estos retos, la profesora Ajunwa ha propuesto la inclusión de una nueva vía de reparación a través del Titulo VII: la discriminación per se. Esta vía de reclamación tomaría como punto de partida las dificultades probatorias a las que se enfrentan los demandantes específicamente cuando se trata de demandar la utilización de sistemas automatizados de selección de personal con resultados discriminatorios. Bajo esta teoría, "un demandante podría afirmar que una práctica de contratación (por ejemplo, el uso de variables proxy en la contratación automatizada que tenga un impacto adverso o el potencial de dar lugar a uno, en las categorías protegidas) es tan notoria que equivale a una discriminación per se, y esto invertiría la carga de la prueba del demandante al empleador para demostrar que su práctica no es discriminatoria"[51]. El mecanismo de la inversión de la carga de la prueba al demandado cumple el objetivo de reducir las barreras procesales que en ocasiones resultan imposibles de superar para los demandantes.

Complementariamente se han sugerido mecanismos específicamente propuestos para prevenir el impacto discriminatorio del uso de algoritmos tales como la implementación de auditorías internas y externas, por evaluadores independientes, que puedan hacer una veeduría de la información de los candidatos que utiliza la herramienta para hacer la preselección, así como qué criterios se utilizan para determinar la habilidad en el empleo entre otra información relevante[52]. Otros mecanismos como la negociación colectiva al

---

afirmación. Rachel Goodman, "Why Amazon's Automated Hiring Tool Discriminated Against Women", American Civil Liberties Union, citado en Ifeoma Ajunwa, "Automated Employment Discrimination", 15.

[50] Sandra F Sperino, "Disparate Impact or Negative Impact? The Future of Non-Intentional Discrimination Claims Brought by the Elderly", *Elder Law Journal* 13, no. 2 (2005): 361.

[51] Ifeoma Ajunwa, "The Paradox of Automation as Anti-Bias Intervention" *Cardozo Law Review* 14 (forthcoming): 45.

[52] Ver, entre otros: Pauline T. Kim, "Auditing Algorithms for Discrimination", *University of Pennsylvania Law Review* 166 (2017): 189-203; Julie E. Cohen, "The Regulatory State in the Information Age", *Theoretical Inquiries* 17 (2016); Ifeoma Ajunwa, "Automated Employment Discrimination", 23.

interior de las empresas podrían promover diseños automatizados justos de selección de personal que incluyan la retención de los datos de los aspirantes (para facilitar la evaluación de universos de candidatos y poder compararlos con el conjunto de seleccionados), así como la adopción de estándares probatorios para determinar si un criterio de selección realmente está relacionado con la labor a desempeñar y que no se trate simplemente de un criterio que sirve para seleccionar adversamente en contra de poblaciones protegidas[53]. Finalmente, también existe una preocupación por el uso de datos que recogen y procesan los mecanismos automatizados de selección. En efecto, los resultados de los análisis en la selección de personal podrían crear "retratos imborrables de los candidatos", lo que puede ser utilizado en el futuro para excluirlos de otras oportunidades de empleo, funcionando en la práctica como un sistema automatizado de balota negra[54]. Por esto, sería importante regular la portabilidad de los datos que entregan los candidatos a un empleo.

Uno de los efectos de las dificultades que hemos mencionado puede ser el relativamente bajo número de casos en Estados Unidos a causa del uso de algoritmos en la selección de personal. No se trata de que los mismos no estén en uso, sino de las amplias dificultades que tienen los trabajadores, bajo el régimen actual, para cumplir con su carga probatoria de manera eficaz, lo que puede funcionar como un desincentivo a demandar.

**6. El vacío regulatorio sobre la discriminación en el acceso al empleo en Colombia**

El sistema jurídico colombiano no cuenta con un estatuto antidiscriminación, y mucho menos con uno especializado en la prevención y sanción de la discriminación en el acceso al empleo. Con lo que contamos es con algunas normas antidiscriminación dispersas en distintas áreas (laboral, penal, administrativo y constitucional) de distintos niveles normativos (desde constitucionales hasta resoluciones ministeriales) que por su carácter desperdigado no comparten métodos de análisis, procedimientos, ni regímenes sancionatorios. La mayoría de estas normas incluyen un compromiso genérico del Estado colombiano con la no discriminación como resultado de la firma de convenios y tratados internacionales, pero pocas de ellas establecen procedimientos administrativos o judiciales para investigar la discriminación como tampoco la imposición de multas o sanciones cuando se incurre en conductas discriminatorias, con excepción de la penalización de actos discriminatorios adoptada en 2015[55]. En el caso particular del Derecho Laboral, las normas existentes prohíben la discriminación *durante* la relación laboral pero no la que ocurre en el período *previo* a la contratación.

---

[53] Ajunwa, "Automated Employment Discrimination", 30

[54] Richard A. Bales y Katherine V. W. Stone, "The Invisible Web of Work: The Intertwining of AI, Electronic Surveillance and Labor Law", *Berkley Journal of Labor and Employment Law* (forthcoming 2020), citado en Ifeoma Ajunwa, "Automated Employment Discrimination", 31.

En efecto, la única norma dentro del título "Principios Generales" del Código Sustantivo del Trabajo que hace alusión a provisiones antidiscriminación es el artículo 10, que fue modificado por la Ley 1496/11 en la cual se adoptaron medidas para prohibir la discriminación salarial entre hombres y mujeres; es decir, el único artículo con un mandato antidiscriminación fue adoptado como parte de una ley sobre discriminación salarial[56]. Adicionalmente, ninguna de las demás normas contenidas en los principios incluye cláusulas antidiscriminación, mientras que el artículo 3 del mismo código precisa que su ámbito de aplicación son las "relaciones de derecho individual del trabajo de carácter particular y las de derecho colectivo del trabajo, oficiales y particulares". Es decir, el código sólo regula las relaciones laborales que ya han nacido a la vida jurídica bien por la firma de un contrato laboral o por la declaración judicial de un contrato realidad. Con base en una interpretación literal de sus competencias, la discriminación en el acceso al empleo estaría fuera del alcance de las labores de vigilancia y control del Ministerio y también excedería las competencias de los jueces laborales[57]. Este vacío jurídico implica que los trabajadores no cuentan con una acción autónoma en la jurisdicción ordinaria laboral a través de la cual resistir la discriminación en el acceso al empleo.

---

[55] El articulo 134A, adoptado por la ley 1752 de 2015, establece como tipo penal, los actos discriminatorios sancionándolos con penas de prisión de 12 a 36 meses y multas de entre diez y quince salarios mínimos legales mensuales.

[56] Artículo 10. Igualdad de los trabajadores y las trabajadoras. Todos los trabajadores y trabajadoras son iguales ante la ley, tienen la misma protección y garantías, en consecuencia, queda abolido cualquier tipo de distinción por razón del carácter intelectual o material de la labor, su forma o retribución, el género o sexo salvo las excepciones establecidas por la ley.

[57] El artículo 2 del Código Procesal del Trabajo y la Seguridad Social indica que la jurisdicción ordinaria conoce de: "1. Los conflictos jurídicos que se originen directa o indirectamente en el contrato de trabajo".

Tal vez la norma que más claramente establece un derecho contra la discriminación y una vía de acción eficaz es el artículo 13 constitucional a través de una acción de tutela. Aún así, existen muy pocos casos en los que la jurisprudencia de la Corte Constitucional haya examinado el tema de la no discriminación en el acceso al empleo. En una investigación que viene realizando el Semillero de Investigación Trabajo y Derecho de la Universidad de los Andes sobre este tema, hemos hecho una revisión de la jurisprudencia constitucional y ordinaria laboral sobre la discriminación en el acceso al empleo en los siguientes tres temas: discriminación de género, racial y por contar con antecedentes penales. Adicionalmente revisamos la regulación de los exámenes pre-ocupacionales, visitas domiciliarias y la aplicación de pruebas de polígrafo como requisitos para acceder a un empleo en Colombia. Nuestra revisión ha encontrado menos de una docena de sentencias en el período 1993-2020 que resuelvan casos relacionados con estos factores pese a que la investigación en ciencias sociales que también hemos revisado en el marco de ese proyecto, da cuenta de un efecto sistemático y estructural de estos factores en las posibilidades de acceder a un empleo para miembros de grupos tradicionalmente discriminados.

La baja litigiosidad que hemos identificado puede ser el resultado de varios factores actuando simultáneamente y con distinta intensidad. Puede ser el caso, por ejemplo, que los candidatos a un empleo no sean conscientes de la discriminación que ocurre cuando en avisos de empleo se especifican categorías prohibidas (hombres y/o mujeres no mayores de 40 años) y por tanto no activen un mecanismo jurídico de protección. También puede ocurrir que, incluso conociendo sus derechos y la posible infracción de los mismos por parte de un posible empleador, los candidatos desestimen el potencial operativo del sistema jurídico para resolver a su favor estas cuestiones, o que los costos económicos y de tiempo superen las expectativas de éxito frente a un eventual litigio. A lo anterior se añade la poca o nula transparencia en los procesos de selección de personal en el sector privado, lo que les impide a los candidatos a un empleo tener información suficiente para notar cuándo un empleador ha tenido en cuenta un criterio discriminatorio para evitar la contratación. Dado que la discriminación en el acceso al empleo es casi una característica de nuestro mercado de trabajo[58] que, como hemos señalado, solo puede agravarse con la utilización de herramientas automatizadas de selección de personal, la regulación del período precontractual en el área laboral es urgente.

**6. Conclusiones**

Formular las bases de un estatuto antidiscriminación supera las posibilidades de este artículo, pero la investigación que hemos presentado hasta aquí podría sugerir algunos elementos a tener en cuenta en el diseño de un estatuto que ofrezca garantías frente a la prevención de la discriminación algorítmica en el acceso al empleo. Recogiendo algunas de las propuestas que presentamos en las secciones anteriores, un estatuto de esta naturaleza debería incluir, como en el régimen estadounidense de impacto diferencial sistémico, la posibilidad de demandar el carácter discriminatorio de una medida con base en sus efectos, excluyendo la necesidad de probar que exista un ánimo discriminatorio por parte del empleador. Al igual que en el contexto estadounidense, las compañías deberían ser responsables de sus decisiones de contratación de personal y por lo tanto deberían ser también responsables de los mecanismos que utilizan para llegar a ellas. A su vez, este régimen antidiscriminación debería reflejar la desigualdad probatoria en la que se encuentran los candidatos demandantes en procesos de selección posiblemente sesgados. Esta desigualdad frente al acceso a la información, particularmente el acceso a la información de los candidatos al empleo y al dominio sobre el algoritmo, debería hacer recaer la carga probatoria de demostrar el carácter no sesgado en la selección de personal en el empleador demandado. Adicionalmente, y con el propósito de facilitar los procesos de auditoría de estos sistemas y evitar por su vía la violación del derecho a la igualdad y al acceso al empleo, la propiedad intelectual sobre estos desarrollos debería ser flexible y admitir tanto auditorías internas como externas del funcionamiento de los algoritmos.

---

[58] Un grupo de estudios recientes ha explorado el efecto de distintos factores de discriminación en el acceso al empleo en Colombia. Con respecto a la discriminación por ubicación de la vivienda del candidato, ver: Ana María Díaz y Luz Magdalena Salas, "Do Firms Redline Workers?", *Regional Science and Urban Economics* 83 (2020): 1-14. Sobre discriminación estadística contra las trabajadoras a causa de la ampliación de la licencia de maternidad, ver: Ana María Tribín, Carmiña Vargas y Natalia Ramírez, "Unintended Consequences of Maternity Leave legislation: The Case of Colombia", *World Development* 122 (2019): 218-232. Sobre la discriminación por maternidad en el acceso al empleo formal, ver: Natalia Ramírez Bustamante, "'A mí me gustaría pero en mis condiciones no puedo'. Maternidad, discriminación y exclusión: el paso del trabajo formal al trabajo informal en confección en Colombia", *Revista CS* (2019): 241-269. Sobre el efecto de la raza del candidato en sus posibilidades de empleo, ver: César Rodríguez Garavito, Juan Camilo Cárdenas, Juan David Oviedo, Sebastián Villamizar, "La discriminación racial en el trabajo. Un estudio experimental en Bogotá", *Documentos Dejusticia* 7 (2013).

Durante los últimos años ha surgido una preocupación generalizada por la falta de transparencia de las decisiones algorítmicas[59]. El Reglamento Europeo de Protección de Datos (GDPR) incluso ha promulgado el "derecho a una explicación"[60] por parte de los usuarios afectados por una decisión algorítmica y el derecho a recibir información significativa acerca la lógica utilizada en esa decisión. La búsqueda de transparencia algorítmica ha llevado a un interés creciente en la IA explicable, (XAI, por sus siglas en inglés), pero todavía es demasiado pronto para saber cuáles métodos serán más efectivos en la tarea de hacer comprensibles las decisiones algorítmicas y si estos métodos lograrán generar confianza por parte de los usuarios[61]. En cualquier caso, es deseable propender por una mayor transparencia algorítmica, al menos en procesos como la selección laboral, que tienen una tendencia a ser discriminatorios.

Por último, aunque en este artículo nos hemos concentrado en la discriminación en el acceso al empleo a través del uso de herramientas de inteligencia artificial, este es sólo uno de los rasgos problemáticos de la aplicación de estas tecnologías en el mundo del trabajo. A medida que la tecnología avanza también se multiplican sus usos en la relación trabajador-empleado. Por ejemplo, con el objetivo de disminuir los costos en salud de sus trabajadores, empresas como Wallmart han contratado los servicios de Castlight Healthcare Inc., para diseñar una aplicación que recoja información sobre la salud de los trabajadores. A través del algoritmo, la aplicación podría identificar quiénes de ellos estaban en riesgo de sufrir enfermedades crónicas tales como diabetes, y enviarles mensajes personalizados para que asistieran a chequeos médicos o se inscribieran en programas de pérdida de peso. Parece un buen objetivo. Sin embargo, los riesgos de privacidad son altos y los expertos advierten que esta información puede ser utilizada para tomar decisiones que afecten la estabilidad en el empleo de los trabajadores.[62] En otro servicio innovador, Castlight lanzó al mercado un servicio para determinar qué trabajadoras podrían quedar en embarazo. En este caso, el algoritmo revisaría las prescripciones médicas de las trabajadoras para identificar a quienes han dejado de reclamar pastillas anticonceptivas y a mujeres que han hecho búsquedas relacionadas con temas de fertilidad en la aplicación. De acuerdo con la empresa, el objetivo es invitar a la trabajadora a elegir a un médico obstetra u otros servicios de cuidado prenatal[63]. Parece una medida bienintencionada, aunque en exceso paternalista. Como en el caso anterior, y dada la fuerte discriminación contra las trabajadoras en embarazo que ha encontrado la investigación sobre esta materia, esta información podría ser utilizada para tomar decisiones sobre la terminación anticipada de un contrato de trabajo fuera del periodo legalmente protegido, ¡incluso antes de que la trabajadora esté en embarazo, o cuando apenas lo está considerando! Estos también son vacíos normativos que necesitamos atender para lograr una buena articulación entre el uso de la tecnología y los derechos de los trabajadores.

---

[59] René Urueña, "Autoridad Algorítmica: ¿Cómo Empezar a Pensar la Protección de los Derechos Humanos en la Era del 'Big Data'?" *Latin American Law Review* 2 (2019): 99-124.

[60] Margot E. Kaminski, "The Right to Explanation, Explained", *Berkeley Technology Law Journal* 34 (2018): 189.

[61] Andrés Páez, "The Pragmatic Turn in Explainable Artificial Intelligence (XAI)", *Minds and Machines* 29, no. 3 (2019): 441-459.

**Bibliografía**

Ajunwa, Ifeoma. "The Paradox of Automation as Anti-Bias Intervention". *Cardozo Law Review* 41 (forthcoming 2020).

Ajunwa, Ifeoma. "Automated Employment Discrimination". *Harvard Journal of Law and Technology* 34 (forthcoming 2021).

Ali, Muhammad, Piotr Sapiezynski, Miranda Bogen, Aleksandra Korolova, Alan Mislove y Aaron Rieke. "Discrimination through Optimization: How Facebook's Ad Delivery Can Lead to Skewed Outcomes". *arXiv preprint arXiv:1904.02095* (2019).

---

[62] Rachel Emma Silverman, "Bosses Tap Outside Firms to Predict Which Workers Might Get Sick", *The Wall Street Journal*, febrero 17, 2016, https://www.wsj.com/articles/bosses-harness-big-data-to-predict-which-workers-might-get-sick-1455664940?st=f4l1g851k9evrdn&reflink=share_mobilewebshare.
[63] Ibid.

Angwin, Julia y Surya Mattu. "How We Analyzed Amazon's shopping algorithm". *ProPublica*, septiembre 20, 2016. https://www.propublica.org/article/how-we-analyzed-amazons-shopping-algorithm.

Barocas, Solon y Andrew D. Selbst. "Big data's disparate impact." *California Law Review* 104 (2016): 671.

Barrett, Lisa Feldman, Ralph Adolphs, Stacy Marsella, Aleix M. Martinez y Seth D. Pollak. "Emotional Expressions Reconsidered: Challenges to Inferring Emotion from Human Facial Movements". *Psychological Science in the Public Interest* 20, no. 1 (2019): 1-68.

Chander, Anupam. "The racist algorithm." *Michigan Law Review* 115 (2016): 1023.

Chen, Le, Ruijun Ma, Anikó Hannák y Christo Wilson. "Investigating the Impact of Gender on Rank in Resume Search Engines". En *Annual Conference of the ACM Special Interest Group on Computer Human Interaction,* 1-14. Montreal, Canadá, abril de 2018.

Cheng, Andria. "Amazon's Coronavirus Plan to Test All 840,000 Employees May Pressure Other Companies to Follow Suit". *Forbes*, abril 16, 2020. https://www.forbes.com/sites/andriacheng/2020/04/16/amazons-plan-to-test-all-employees-on-coronavirus-may-up-the-ante-for-other-companies/#b4e0a0736cb1.

Cohen, Julie E. "The Regulatory State in the Information Age". *Theoretical Inquiries in Law* 17, no. 2 (2016): 369.

Dastin, Jeffrey. "Amazon Scraps Secret AI Recruiting Tool that Showed Bias Against Women". *Reuters*, octubre 9, 2018. https://www.reuters.com/article/us-amazon-com-jobs-automation-insight/amazon-scraps-secret-ai-recruiting-tool-that-showed-bias-against-women-idUSKCN1MK08G.

Datta, Amit, Michael Carl Tschantz y Anupam Datta. "Automated Experiments on Ad Privacy Settings: A Tale of Opacity, Choice, and Discrimination". *Proceedings on Privacy Enhancing Technologies* 2015, no. 1 (2015): 92-112.

Díaz, Ana María y Luz Magdalena Salas. "Do Firms Redline Workers?" *Regional Science and Urban Economics* 83 (2020): 1-14.

Dwork, Cynthia, Moritz Hardt, Toniann Pitassi, Omer Reingold y Richard Zemel. "Fairness through Awareness". In *Proceedings of the 3rd Innovations in Theoretical Computer Science Conference*, 214-226. New York: ACM, 2012.

Dwork, Cynthia y Christina Ilvento. "Individual fairness under composition". *Proceedings of Fairness, Accountability, Transparency in Machine Learning*. New York: ACM, 2018.

Edelman, Benjamin, Michael Luca y Dan Svirsky. "Racial Discrimination in the Sharing Economy: Evidence from a Field Experiment". *American Economic Journal: Applied Economics* 9, no. 2 (2017): 1-22.

Facebook (2020). "Advertising Policies: 3. Discriminatory Practices. https://www.facebook.com/policies/ads/prohibited_content/discriminatory_practices#

Hannák, Anikó, Claudia Wagner, David Garcia, Alan Mislove, Markus Strohmaier y Christo Wilson. "Bias in Online Freelance Marketplaces: Evidence from Taskrabbit and Fiverr". En *Proceedings of the 2017 ACM Conference on Computer Supported Cooperative Work and Social Computing*, 1914-1933. New York: ACM, 2017.

Harwell, Drew. "A Face-Scanning Algorithm Increasingly Decides whether You Deserve the Job." *The Washington Post*, octubre 22, 2019. https://www.washingtonpost.com/technology/2019/10/22/ai-hiring-face-scanning-algorithm-increasingly-decides-whether-you-deserve-job/.

Illinois Attorney General. "Madigan Probes National Job Search Sites over Potential Age Discrimination". *Illinois Attorney General Press Release,* March 2, 2017. http://www.illinoisattorneygeneral.gov/pressroom/2017_03/20170302.html.

Kaminski, Margot E. "The Right to Explanation, Explained". *Berkeley Technology Law Journal* 34 (2018): 189.

Kim, Pauline T. "Data-driven discrimination at work". *William & Mary Law Review* 58 (2016): 857.

Kim, Pauline T. "Auditing Algorithms for Discrimination". *University of Pennsylvania Law Review* 166 (2017): 189.

OECD. "Recommendations of the Council on AI". *OECD Legal Instrument 0449* (2019).

Páez, Andrés. "The Pragmatic Turn in Explainable Artificial Intelligence (XAI)". *Minds and Machines* 29, no. 3 (2019): 441-459.

Rainie, Lee, Janna Anderson y D. Page. "Code-Dependent: Pros and Cons of the Algorithm age". *Pew Research Center,* febrero 8, 2017. https://www.pewresearch.org/internet/2017/02/08/code-dependent-pros-and-cons-of-the-algorithm-age/.

Ramírez Bustamante, Natalia. "'A mí me gustaría pero en mis condiciones no puedo'. Maternidad, discriminación y exclusión: el paso del trabajo formal al trabajo informal en confección en Colombia", *Revista CS* (2019): 241-269.

Rodríguez Garavito, César, Juan Camilo Cárdenas, Juan David Oviedo y Sebastián Villamizar. "La discriminación racial en el trabajo. Un estudio experimental en Bogotá". *Documentos Dejusticia* 7 (2013).

Selmi, Michael. "Was the Disparate Impact Theory a Mistake?" *UCLA Law Review* 53 (2005): 701.

Shields, Jon. "Over 98% of Fortune 500 companies use Applicant Tracking Systems (ATS)". *Jobscan*, junio 20, 2018. https://www.jobscan.co/blog/fortune-500-use-applicant-tracking-systems/.

Silverman, Rachel Emma. "Bosses Tap Outside Firms to Predict Which Workers Might Get Sick". *The Wall Street Journal*, febrero 17, 2016. https://www.wsj.com/articles/bosses-harness-big-data-to-predict-which-workers-might-get-sick-1455664940.

Sperino, Sandra F. "Disparate Impact of Negative Impact: Future of Non-Intentional Discrimination Claims Brought by the Elderly." *Elder Law Journal* 13 (2005): 339.

Sweeney, Latanya. "Discrimination in Online Ad Delivery". *Queue* 11, no. 3 (2013): 10-29.

Tribín, Ana María, Carmiña Vargas y Natalia Ramírez. "Unintended Consequences of Maternity Leave Legislation: The Case of Colombia". *World Development* 122 (2019): 218-232.

Urueña, René. "Autoridad Algorítmica: ¿Cómo Empezar a Pensar la Protección de los Derechos Humanos en la Era del 'Big Data'?" *Latin American Law Review* 2 (2019): 99-124.